\documentclass{article}  
\usepackage{graphics, graphicx}          
\usepackage{natbib}
\bibpunct{(}{)}{;}{a}{}{;}
\newcommand{\quot}[1]{`#1'}

\newcommand{\D}[1]{D_{#1}}
\newcommand{\A}{\widehat{A}}

\newcommand{\obsom}{\omega^\star}
\fussy
\begin{document}

\title{Parametric resonance as a model for QPO sources -- I.\\
A general approach to multiple scales\footnote{To
appear in proceedings of the Workshop {\it Processes in the vicinity of 
black holes and neutron stars}, held at Silesian University (Opava, October 
2003).}}

\author{Ji\v{r}\'{\i} Hor\'ak\footnote{Faculty of Mathematics and Physics,
  Charles University in Prague, Czech Republic}}
\maketitle

\begin{abstract}
In the resonance model, high-frequency quasi-periodic oscillations 
(QPOs) are supposed
to be a consequence of nonlinear resonance between modes of
oscillations occurring within the innermost parts of an accretion disk. 
Several models
with a prescribed mode--mode interaction were proposed 
in order to explain the characteristic properties of the 
resonance in QPO sources. In this paper, we 
examine nonlinear oscillations of a system
having a quadratic nonlinearity and we show that this case is
particularly relevant for QPOs. We present a very
convenient way how to study internal resonances of a fully general system
using the method of multiple scales. Finally, we concentrate to
conservative systems and discuss their behavior near the $3:2$ resonance. 
\end{abstract}

%
%

\section{Introduction}
In the resonance model \citep{ak01,ka01,kato03}, there is a
natural and attractive possibility of explaining the observed rational
ratios of high-frequency QPOs as a consequence of non-linear coupling 
between different modes of
accretion disk oscillations. The idea has been pursued in several
papers \citep[recently, e.g.][]{akklr03, r04}. 

Specific models invoke particular physical mechanisms. Some models
can be almost immediately comprehended as
distinct realisations of the general approach discussed here -- for example,
various formulations of the orbiting spot model \citep{sb04}
or the models, where QPOs are produced by the magnetically driven resonance 
in a diamagnetic accretion disk \citep{lai99} -- while other seem to be
more distant from the view presented herein -- e.g.\ the transition layer
model \citep{tit02}, an interesting idea of
p-mode oscillations of a small accretion torus \citep{rezzola03}
or the model of blobs in an accretion disc \citep[see e.g.][and references
cited therein]{ka99,li04}. 
Also in this context, \citet{kato04} discussed the resonant interaction 
between waves propagating in a warped disk, including their rigorous mathematical
description.
Instead of pursuing a specific model, here we keep the discussion as
general as possible, aiming to implement the formalism of multiple scales.
Indeed, we show that there is unquestionable appeal in this approach which
offers some additional insight into generic properties of resonant
oscillations.

Some properties of an accretion disk oscillations can be 
discussed within the epicyclic approximation of a
test particle on a circular orbit near equatorial plane. 
Suppose that
angular momentum of the particle is fixed to a value $ \ell $. The
effective potential $ U_\ell(r, \theta) $ has a minimum at radius $ r_0
$, corresponding to the location of the stable circular orbit. An observer
moving along this orbit measures radial, vertical and azimuthal
epicyclic oscillations of a particle nearby. Since the angular momentum of
the particle is conserved, only two of them -- radial and vertical --
are independent. The epicyclic frequencies can be derived from the
geodesic equations expanded to the linear order in deviations $ \delta r
= r - r_0 $ and $ \delta \theta = \theta - \pi/2 $ from the circular
orbit. We get two independent second-order differential equations
describing two uncoupled oscillators with frequencies $ \omega_r $ and $
\omega_\theta $, which are given by the second derivatives of effective
potential $ U_\ell(r, \theta) $. In Newtonian theory, $ \omega_r $ and $
\omega_\theta $ are equal to the Keplerian orbital frequency $ \Omega_K
$. This is in tune with the fact that orbits of particles are planar and
closed curves. The degeneracy between two epicyclic frequencies can be
seen as a result of scale-freedom of the Newtonian gravitational
potential \citep{ak03}. In Schwarzschild geometry this freedom is broken
by introducing the gravitational radius $ r_g = 2GM/c^2 $. The
degeneracy between the vertical epicyclic and the orbital frequencies is
related to spherical symmetry of the gravitational potential, which
assures the existence of planar trajectories of particles. All three
frequencies are different in the vicinity of a rotating Kerr black hole.

In addition, when nonlinear terms of geodesic equations are included, the 
two oscillations in $ r $ and $\theta $ directions become coupled and 
variety of new phenomena connected to nonlinear nature of the equations 
appear. This rich phenomenology includes frequency shift of observed frequencies 
with respect to eigenfrequencies, presence of higher harmonics and 
subharmonics, drifts and parametric resonance. The first 
three are connected to nonlinear oscillations of each mode and the last one 
comes from the coupling between two modes.

The paper is organized in a following way: In the next section we will study 
one-degree-of-freedom system with a quadratic nonlinearity. We introduce 
the method of multiple scales \citep{n73, nm79} and, using this formalism, 
we derive key properties of nonlinear oscillations. In the third section we perform the multiple-scales 
expansion in the case of a fully general system of two coupled oscillators 
and find possible parametric resonances up to the fourth order. Then we 
concentrate on the $3:2$ parametric resonance of the conservative system. 

%
%

\section{Effects of nonlinearities}

\label{sec:qn}

Let us consider the case of small but finite
oscillations of a single-degree-of-freedom system with quadratic
nonlinearity governed by equation 
\begin{equation}
\ddot{x} + \omega^2 x = \alpha \omega^2 x^2.
\label{eq:qn_gov}
\end{equation}
The strength of the nonlinearity is parametrized by constant $ \alpha $.
When $ \alpha = 0 $ one obtains governing equation of the corresponding
linear system. We seek a perturbation expansion of the form 
\begin{equation}
x(t, \epsilon) = \epsilon x_1(t) + \epsilon^2 x_2(t) + \epsilon^3 x_3(t) + {\cal O}(\epsilon^4),
\label{eq:qn_exp}
\end{equation}
The expansion parameter $ \epsilon $ expresses the order of amplitude of
oscillations. The main advantage of this approach is that, although the
original equation is nonlinear, we solve linear equations in each step. 
For a practical purpose we require this
expansion to be uniformly convergent for all times of interest. In that
case the higher-order terms are small compared to lower-order terms and
a sufficient approximation is reached concerning the finite number of
terms. The expansion (\ref{eq:qn_exp}) can represent
a periodic solution as well as an unbounded solution with exponential grow.
The uniformity of the expansion means that the higher-order
terms are not larger than the corresponding lower-order ones.

We substitute expansion (\ref{eq:qn_exp}) into governing equation
(\ref{eq:qn_gov}) and, since $ x_k $ is independent of $ \epsilon $,  we
equate coefficients of corresponding powers of $ \epsilon $ on both
sides. This leads to the following system of equations
\begin{eqnarray}
\label{eq:qn_1}
\ddot{x}_1 + \omega^2 x_1 &=& 0, \\
\label{eq:qn_2}
\ddot{x}_2 + \omega^2 x_2 &=& \alpha \omega^2 x_1^2, \\
\label{eq:qn_3}
\ddot{x}_3 + \omega^2 x_3 &=& 2\alpha \omega^2 x_1 x_2.
\end{eqnarray}
The general solution of eq.\ (\ref{eq:qn_1}) can be written in the form $
x_1(t) = A e^{i \omega t} + \mathrm{cc} $, where $ \mathrm{cc} $ denotes
complex conjugation. The complex constant $ A $ contains information about
the initial amplitude and phase of oscillations. Substituting it into
eq.\ (\ref{eq:qn_2}) we find linear equation for the first approximation $
x_2(t) $ 
\begin{equation}
\label{eq:qn_2b}
\ddot{x}_2 + \omega^2 x_2 = \alpha \left( A^2 e^{2i \omega t} + |A|^2 \right) + \mathrm{cc}.
\end{equation}
A general solution consists of the solution of the homogeneous equation and
a particular solution,
\begin{equation}
x_2(t) = A_2 e^{i \omega t} - \alpha \left( \frac{1}{3} A_1^2 e^{2i \omega t} - |A_1|^2 \right) + \mathrm{cc},
\end{equation}
where $ A_1 $ denotes a constant $ A $ of the solution of
eq.~(\ref{eq:qn_1}). Therefore, the solution of governing equation up to
the second order is given by
\begin{equation}
\label{eq:qn_2gensol}
x(t) = \left(\epsilon A_1 + \epsilon^2 A_2 \right) e^{i \omega t} - \alpha \left( \frac{1}{3} A_1^2 e^{2i \omega t} - |A_1|^2 \right) + \mathrm{cc}.
\end{equation}
In fact, there are two possible ways how to satisfy general initial
conditions $ x(0) = \epsilon x_0 $ and $ \dot{x}(0) = \epsilon \dot{x}_0
$ imposed on equation (\ref{eq:qn_gov}). The first one is to compare them with the
general solution (\ref{eq:qn_2gensol}) and find constants $ A_1 $, $ A_2
$. This procedure should be repeated in each order of approximation
which involves quite complicated algebra especially in higher orders.
The second, equivalent and apparently much easier way is to include only
particular solutions to the higher approximations and treat the constant
$ A $ as a function of $ \epsilon $ with expansion  $ A = A_1 + \epsilon
A_2  + \ldots $. Then, given initial conditions are satisfied by
expanding the solution for $ x_1 $ via $ \epsilon $ and choosing the
coefficients $ A_n $ appropriately. 

According to this discussion we express the solution of eq.~(\ref{eq:qn_2b})
as  
\begin{equation}
\label{eq:qn_x2}
x_2(t) = - \alpha \left( \frac{1}{3} A^2 e^{2i \omega t} - |A|^2 \right) + \mathrm{cc}.
\end{equation}
Substituting $ x_1 $ and $ x_2 $ into (\ref{eq:qn_3}) we obtain
\begin{equation}
\ddot{x}_3 + \omega^2 x_3 = \frac{2\alpha^2\omega^2}{3} \left(  5 A|A|^2 e^{i \omega t} - A^3 e^{3i \omega t} \right) + \mathrm{cc}.
\end{equation}
Since the right-hand side of this equation contains the term proportional to
$ e^{i \omega t} $, any solution must contain a secular term
proportional to $ t e^{i \omega t} $, which becomes unbounded as $ t
\rightarrow \infty $. This fact has nothing to do with true
physical behavior of the system for large times. The meaning is rather
mathematical. Starting from time when $ (\omega t) \sim 1/(\epsilon
\alpha) $, the  higher-order approximation $ x_3 $, which contains the
secular term, does not provide only a small correction to $ x_1 $ and $ x_2 $,
and the expansion (\ref{eq:qn_exp}) becomes singular. However, the presence of
the secular term in the third order reflects very general
feature of nonlinear oscillations -- dependence of the observed
frequency on the actual amplitude. For larger amplitudes the actual
frequency of oscillations differs from the eigenfrequency $ \omega $ and
the higher-order terms in the expansion (\ref{eq:qn_exp}) -- always
oscillating with an integer multiples of $ \omega $ -- must quickly
increase as time grows. 


\subsection{The method of multiple scales}
Is it possible to find an expansion representing a solution of equation
(\ref{eq:qn_gov}) which is uniformly valid even for larger time then $
\sim \epsilon^{-1} $? Yes, if one considers more general
form of the expansion than eq.~(\ref{eq:qn_exp}). In the method of multiple
scales  more general dependence of coefficients $ x_i $ on the time is
reached by introducing several time scales $ T_\mu $, instead of one
physical time $ t $. The time scales are introduced as
\begin{eqnarray}
T_\mu \equiv \epsilon^\mu t, &  \mu = 0,1,2\ldots
\label{eq:ms_scales}
\end{eqnarray}
and they are treated as independent. It follows that instead of the
single time derivative we have an expansion of partial derivatives with
respect to the $ T_\mu $
\begin{eqnarray}
\label{eq:ms_der1}
\frac{d}{dt} &=& \D{0} + \epsilon \D{1} + \epsilon^2 \D{2} + \epsilon^3 \D{3} + 
{\cal O}(\epsilon^4), \\
\label{eq:ms_der2}
\frac{d^2}{dt^2} &=& \D{0}^2 + 2 \epsilon \D{0} \D{1} + 
\epsilon^2 (\D{1}^2 + 2 \D{0}\D{2}) + 2 \epsilon^3 (\D{0}\D{3} + \D{1}\D{2}) + 
{\cal O}(\epsilon^4),
\end{eqnarray}
where $ \D{\mu} = \partial / \partial T_\mu $.

We assume that the solution can be represented by an expansion having
the form
\begin{equation}
x(t, \epsilon) = \epsilon x_1(T_\mu) + \epsilon^2 x_2(T_\mu) + \epsilon^3 x_3 (T_\mu) + {\cal O}(\epsilon^4).
\label{eq:ms_exp}
\end{equation}
The number of time scales is always the same as the order at which the
expansion is truncated. Here we carry out the expansion to the third
order and thus first three scales $ T_0 $, $ T_1 $ and $ T_2 $ are
sufficient.

Substituting eqs.~(\ref{eq:ms_exp}) and (\ref{eq:ms_der2}) into the governing
equation (\ref{eq:qn_gov}) and equating the coefficients of $ \epsilon
$, $ \epsilon^2 $ and $ \epsilon^3 $ to zero we obtain
\begin{eqnarray}
\label{eq:ms_1}
(\D{0}^2 + \omega^2) x_1 &=& 0, \\
\label{eq:ms_2}
(\D{0}^2 + \omega^2) x_2 &=& -2\D{0}\D{1} x_1 + \alpha \omega^2 x_1, \\
\label{eq:ms_3}
(\D{0}^2 + \omega^2) x_3 &=& -2\D{0}\D{1} x_2 - \D{1}^2 x_1 - 2\D{0}\D{2} x_1 + \alpha \omega^2 x_1 x_2.
\end{eqnarray}
Note that although these equations are more complicated than
(\ref{eq:qn_1})--(\ref{eq:qn_3}), they are still linear and can be
solved successively. The solution of the equation (\ref{eq:ms_1}) is the
same as the solution of the corresponding linear system, the only difference
is that constant $ A $ now generally depends on other scales
\begin{equation}
\label{eq:ms_1sol}
x_1 = A(T_1, T_2) e^{i \omega T_0} + \mathrm{cc}.
\end{equation}
Substituting $ x_1 $ into the equation (\ref{eq:ms_2}) we obtain
\begin{equation}
\label{eq:ms_2sub}
(\D{0}^2 + \omega^2) x_2 = -2i\omega (\D{1}A) e^{i \omega T_0} + \alpha
\omega^2 \left( A^2 e^{2i \omega T_0} + |A|^2 \right) + \mathrm{cc}.
\end{equation}
The first term on the right-hand side implies the presence of a 
secular term in the second-order 
approximation which causes non-uniformity of the expansion
(\ref{eq:ms_exp}). However, in case of the method of multiple scales these
terms can
be eliminated by imposing additional conditions on the function $ A(T_\mu) $. 
These conditions are sometimes called conditions of solvability or
conditions of consistency. In fact, the reason why the same number of scales as 
the order of approximation is needed is that one secular term 
gets eliminated in each order, and therefore the function $ A(T_\mu) $ is 
specified by the same number of solvability conditions as 
the number of its variables.

The secular term is eliminated if we require $ \D{1} A = 0 $. 
In further discussion we assume that $ A $ is a function of $ T_2 $ only. 
A particular solution of equation (\ref{eq:ms_2sub}) is
\begin{equation}
\label{eq:ms_2sol}
x_2(t) = - \alpha \left( \frac{1}{3} A^2(T_2) e^{2i \omega T_0} -
|A(T_2)|^2 \right) + \mathrm{cc}.
\end{equation}
Using the condition $ \D{1} A = 0 $ the equation (\ref{eq:ms_3}) takes
much simpler form
\begin{equation}
\label{eq:ms_3sub}
(\D{0}^2 + \omega^2) x_3 = - \left(2i \omega \D{2}A + \frac{10 \alpha^2
\omega^2}{3} A |A|^2 \right) e^{i \omega T_0} - \frac{2 \alpha^2
\omega^2}{3} A^3 e^{3i \omega T_0} + \mathrm{cc}.
\end{equation}
The secular term is eliminated equating the terms in the bracket to
zero
\begin{equation}
\label{eq:ms_3sec}
2i \omega \D{2}A + \frac{10 \alpha^2 \omega^2}{3} A |A|^2 = 0.
\end{equation}
This additional condition fully determines (excepting initial
conditions) time behavior of function $ A(T_2) $. 
Let us write it in the polar form $ A = \frac{1}{2} \tilde{a}
e^{i\phi} $ and then separate real and imaginary parts. We obtain
\begin{eqnarray}
\D{2}\tilde{a} = 0 & \mathrm{and} & \D{2}\phi = - \frac{5 \alpha^2}{12}
\omega \tilde{a}^2.
\end{eqnarray}
The solutions of these equations are
\begin{eqnarray}
\label{eq:ms_Asol}
\tilde{a} = \tilde{a}_0
& \mathrm{and} &
\phi = -\frac{5}{12}\alpha^2 \tilde{a}_0^2 \omega T_2 + \phi_0,
\end{eqnarray}
where $ \tilde{a}_0 $ and $ \phi_0 $ are constants which are determined
from the initial conditions. 

It follows from eq.\ (\ref{eq:ms_1sol}) that $ A(T_2) $ slowly modulates the
amplitude and the phase of oscillations. Since $ \tilde{a} $ is
constant, the amplitude is constant all the time. Since $ \phi $ depends
on $ T_2 = \epsilon^2 t $ linearly, also the observed frequency of the
oscillations is constant, but not equal to the eigenfrequency $ \omega $.

Substituting eqs.~(\ref{eq:ms_Asol}) and (\ref{eq:ms_2sol}) into
(\ref{eq:ms_exp}), we obtain solution of eq.~(\ref{eq:qn_gov}) up to the
second order
\begin{equation}
\label{eq:ms_sol}
x(t) = a_0 \cos(\obsom t + \phi_0) - \frac{\alpha}{6} a_0^2
\cos[2(\obsom t + \phi_0)] + \frac{\alpha}{2} a_0^2 + {\cal O}(a_0^3),
\end{equation}
where $ a_0 = \epsilon \tilde{a}_0 \ll 1 $ and $ \obsom $ is the
observed frequency of oscillation given by
\begin{equation}
\label{eq:ms_obsom}
\obsom = \omega \left( 1 - \frac{5 \alpha^2}{12} a_0^2 \right).
\end{equation}


\subsection{Essential properties of nonlinear oscillations}

\begin{figure}[t!]
\begin{center}
\includegraphics[scale=1.4]{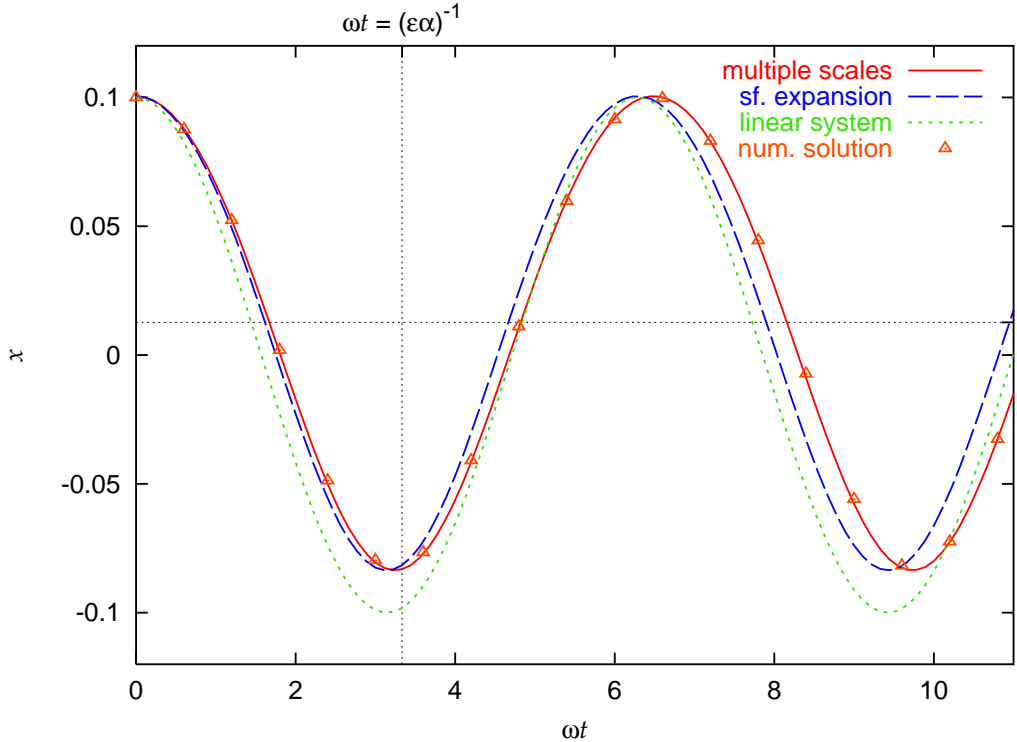}
\end{center}
\caption{Oscillations of the system with quadratic nonlinearity governed
by the equation (\ref{eq:qn_gov}). We compare results of the multiple
scales method (solid curve), a simple straightforward expansion method
(dashed curve) and the direct numerical integration of the governing 
equation (points). The
initial condition is $ x(0) = 0.1 $, $ \dot{x}(0) = 0 $ and the
strength of nonlinearity is $ \alpha = 3 $. The horizontal dotted line
shows the shifted \quot{equilibrium position} and the vertical one
denotes the value $ (\omega t) = (\epsilon \alpha)^{-1} $ at which the
straightforward expansion becomes nonuniform. The 
solution corresponding to the linear system is also
shown (dotted curve).}
\label{fig:methods}
\end{figure}

We close this section summarizing main properties of nonlinear
oscillations. The equation (\ref{eq:ms_sol}) is very helpful for this
purpose.
  
The leading term of the expansion (\ref{eq:ms_sol}) describes
oscillations with frequency close to eigenfrequency of the system. Both
the amplitude and the frequency are constant in time, but they are not
independent (as in case of the linear approximation). The frequency
correction given by (\ref{eq:ms_obsom}) is proportional to the square of
the amplitude. This fact is sometimes called the amplitude-frequency
interaction and -- as was mentioned above -- it causes the non-uniformity of
the expansion (\ref{eq:qn_exp}) (see the figure \ref{fig:methods}). 

The second term oscillates with twice as large frequency and provides the
second-order correction to the leading term. The presence of higher
harmonics is another particular feature of nonlinear oscillations and
the fact that it has been reported in several sources of QPOs, points to
nonlinear nature of this phenomena.

The third term describes constant shift from the equilibrium position
and is related to the asymmetry of the potential energy about the point $ x
= 0 $. In the linear analysis, this effect is not present because the potential
energy depends on $ x^2 $, and therefore is symmetric with respect 
to the point $ x=0 $. Hence, the drift is the third characteristic 
feature of nonlinear oscillations. 

%
%

\section{Parametric resonances in conservative systems}

\label{sec:res}

Let us study nonlinear oscillations of the system having two degrees
of freedom, i.e., the coordinate perturbations $ \delta r $ and 
$ \delta \theta $. The oscillations are
described by two coupled differential equations of the very general form
\begin{eqnarray}
\label{eq:res_gov_r}
\ddot{\delta r} + \omega_r^2\; \delta r &=& \omega_r^2\; f_r(\delta r,
\delta\theta, \dot{\delta r},\dot{\delta \theta}), 
\\
\label{eq:res_gov_theta}
\ddot{\delta \theta} + \omega_\theta^2\; \delta \theta &=& \omega_\theta^2
\;f_\theta(\delta r, \delta\theta, \dot{\delta r}, \dot{\delta \theta}).
\end{eqnarray}
Suppose that the functions $ f_r $ and $ f_\theta $ are nonlinear, i.e.,
their Taylor expansions start in the second order. Another assumption is
that these functions are invariant under reflection of time (i.e.,
the Taylor expansion does not contain odd powers of time derivatives of
$ \delta r $ and $ \delta \theta $). As we see later, this assumption 
is related to the
conservation of energy in the system. Many authors studied such systems
with a particular form of functions $ f $ and $ g $ \citep{nm79}, however, in this
paper we keep discussion fully general. 

We seek the solutions of the governing equations in the form of the
multiple-scales expansions
\begin{eqnarray}
\label{eq:res_exp_r}
\delta r(t, \epsilon) = \sum_{n=1}^4 \epsilon^n r_n(T_\mu),
&&
\label{eq:res_exp_m}
\delta \theta(t, \epsilon) = \sum_{n=1}^4 \epsilon^n \theta_n(T_\mu),
\end{eqnarray}
where $ T_\mu = \epsilon^\mu t $ are independent time scales, $ \mu =
0,1,2,3 $ (we finish the discussion in the fourth order, however, it
is possible to proceed to higher orders in suggested way). We expand the
time derivatives according to eqs.\ (\ref{eq:ms_der1}) and (\ref{eq:ms_der2})
and equate terms of the same order in $ \epsilon $ on both sides of
the governing equations. 

In the first order we obtain equations corresponding to the linear
approximation
\begin{eqnarray}
\label{eq:res_1}
(\D{0}^2 + \omega_r^2) r_1 = 0,
&&
(\D{0}^2 + \omega_\theta^2) \theta_1 = 0.
\end{eqnarray}
with the solutions
\begin{eqnarray}
\label{eq:res_1sol}
x_1 = \A_r + \A_{-r},
&&
\theta_1 = \A_\theta + \A_{-\theta},
\end{eqnarray}
where we denoted $ \A_x = A_x e^{i \omega_x T_0} $, $ \A_{-x} =
A_x^\ast e^{-i \omega_x T_0} $, and $ x = r$, $\theta $, respectively 
(since many considerations are independent of the
mode of oscillations, we keep this notation through the whole section). 
The complex functions $ \A_x $ generally
depend on the time-scales $ T_1 $, $ T_2 $, $ T_3 $. 

Having solved the linear approximation, we can proceed to higher orders.
The terms proportional to $ \epsilon^2 $ in the expanded left-hand side
of the governing equation (\ref{eq:res_gov_r}) resp.
(\ref{eq:res_gov_theta}) are
\begin{equation}
\label{eq:res_2lhs}
\left[ \ddot{\delta x} + \omega_x^2 x \right]_2 = (\D{0}^2 + \omega_x^2)
x_2  +  2i\omega_x\D{1}\A_x  -  2i\omega_x\D{1}\A_{-x},
\end{equation}
On the right-hand side there are second-order terms of the Taylor
expansion of the nonlinearity $ f(\delta r, \delta\theta, \dot{\delta
r}, \dot{\delta \theta}) $, with $ r_1 $, $ \theta_1 $, $ \D{0} r_1 $
and $ \D{0}\theta_1 $ in the place of $ \delta r $, $ \delta \theta $, $
\dot{\delta r} $ and $ \dot{\delta \theta} $, respectively. 
Since the derivative with respect to $T_0$ 
only adds the coefficient $ i \omega_x$, the second-order terms 
on the right-hand sides 
can be expressed as linear combinations of quadratic terms
constructed from $ \A_{\pm r} $ and $ \A_{\pm \theta} $
\begin{equation}
\label{eq:res_2rhs}
\left[ f_x(\delta r, \delta\theta, \dot{\delta r}, \dot{\delta \theta})
\right]_2 = \omega_x^2 \sum_{|\alpha|=2} C_\alpha^{(2,x)}
\A_r^{\alpha_1} \A_\theta^{\alpha_2} \A_{-r}^{\alpha_3}
\A_{-\theta}^{\alpha_4},
\end{equation}
where $ \alpha = (\alpha_1, \ldots, \alpha_4) $ and $ |\alpha| =
\alpha_1 + \ldots + \alpha_4 $. The constants $ C_\alpha^{(2,x)} $ are
combinations of $\omega_x$ and the coefficients in the Taylor expansion 
of functions $f$ and $g$, respectively.
The coefficients coming from the terms containing time derivatives
are generally complex, since each time derivative produces one ``$i$''.
However, if we suppose that the Taylor expansion does not contain odd
powers of time derivatives, all of the constants $ C_\alpha^{(2,x)} $
must be real. Equating right-hand sides of equations 
(\ref{eq:res_2lhs}) and (\ref{eq:res_2rhs}) we have
\begin{equation}
\label{eq:res_2}
(\D{0}^2 + \omega_x^2) x_2 = - 2i\omega_x \D{1}\A_x  + 
2i\omega_x\D{1}\A_{-x}  +  \omega_x^2 \sum_{|\alpha|=2} C_\alpha^{(2,x)}
\A_r^{\alpha_1} \A_\theta^{\alpha_2} \A_{-r}^{\alpha_3}
\A_{-\theta}^{\alpha_4}.
\end{equation}

%
%

\begin{table}[t]
\begin{center}
\begin{tabular}{c@{~~~}c@{~~~}r}
\hline
\rule{0mm}{3.5ex} \rule[-3ex]{0mm}{3.5ex}
$ \omega_\theta : \omega_r $  &
\textbf{Secular terms}  &
\textbf{Solvability condition~~} \\
\hline\hline
\rule{0mm}{3.5ex} \textrm{Outside} &
$\D{1}\A_r $&
$\D{1}\A_r = 0$
\\
 \textrm{resonance} &\rule[-2.5ex]{0mm}{3.5ex}
 $\D{1}\A_\theta$ &
 $\D{1}\A_\theta = 0$
 \\
 \hline
 \rule{0mm}{3.5ex}
 $1:2$ &
 $\A_{-r}\A_\theta$ &
 $-2i\D{1}\A_r + \omega_r K\A_{-r}\A_\theta = 0$
 \\
  ~ &
  $\A_r^2$ &
  $-2i\D{1}\A_\theta + \omega_\theta L\A_r^2 = 0$\rule[-2.5ex]{0mm}{3.5ex}
  \\
  \hline
  \rule{0mm}{3.5ex}
  $2:1$ &
  $\A_\theta^2$ &
  $-2i\D{1}\A_r + \omega_r K\A_\theta^2 = 0$
  \\
   &
   $\A_{r}\A_{-\theta}$ &
   $-2i\D{1}\A_\theta + L\A_r A_{-\theta} = 0$\rule[-2.5ex]{0mm}{3.5ex}
   \\
\hline
\end{tabular}
\caption{Possible resonances and appropriate solvability conditions in
the second order of approximation. We substitute constants $
C^{(n,x)}_\alpha $ by $ K $ and $ L $ for simplicity. The first record
is related to the case when the system is far from any listed resonance.
In this case only strictly secular terms are present. The first resp.
second row in the record of each resonance is related to the equation
for the radial resp. vertical oscillations. In that case, we list only
nearly secular terms in the 2nd column, however strictly secular terms
are included in the solvability conditions.}
\label{tab:res_2}
\end{center}
\end{table}

The right-hand side of equation (\ref{eq:res_2}) contains one
secular term independently of the eigenfrequencies $ \omega_r $ and $
\omega_\theta $. However, in some particular cases, 
additional secular terms appear. For example, when $
\omega_r \approx 2 \omega_\theta $ the terms proportional to $
\A^2_\theta $ in the radial equation ($x\rightarrow r$) and 
$ \A_r \A_{-\theta} $ in the
meridional equation ($x\rightarrow\theta$) become nearly secular 
and they should be included
in the solvability conditions. The analogical situation happens when $
\omega_r \approx  \omega_\theta / 2 $. These cases
show qualitatively different behavior, and we speak about
parametric or internal resonance. Possible resonances
in the second order of approximation and appropriate solvability
conditions are listed in the table \ref{tab:res_2}. At this moment, let
us assume that the system is far from the resonance and require
\begin{equation}
\label{eq:res_2sec}
\D{1} A_x = 0.  
\end{equation}
In this case the frequencies and the amplitudes are constant
and the behavior of the system is almost the same as in the linear
approximation. The only difference is the presence of the higher
harmonics oscillating with the frequencies $ 2 \omega_r $, $ 2
\omega_\theta $ and $ |\omega_r \pm \omega_\theta| $. They are given by
a particular solution of equation (\ref{eq:res_2}) after elimination of
secular term and can be expressed as a linear combination
\begin{equation}
\label{eq:res_2sol}
x_2 = \sum_{|\alpha|=2} Q_\alpha^{(2,x)} \A_r^{\alpha_1}
\A_\theta^{\alpha_2} \A_{-r}^{\alpha_3} \A_{-\theta}^{\alpha_4}.
\end{equation}
Under the assumption of invariance under the time reflection,
constants $ Q_\alpha^{(2,x)} $ are real and their relation to $
C_\alpha^{(2,x)} $ becomes obvious, by substituting $ x_2 $ into equation 
(\ref{eq:res_2}).

When we proceed to the higher order, the discussion is analogical in
many respects. The terms proportional to $ \epsilon^3 $, which appear on
the left-hand side of the governing equations, are given by 
\begin{equation}
\left[ \ddot{\delta x} + \omega_x^2 x \right]_3 =
(\D{0}^2 + \omega_x^2) x_3  +  2i\omega_x\D{2}\A_x  - 
2i\omega_x\D{2}\A_{-x}.
\end{equation}
The terms containing $ D_1 x_1 $ and $ D_1 x_2 $ vanish in consequence
of the solvability condition (\ref{eq:res_2sec}). The right-hand side
contains cubic terms of the Taylor expansion combined using first-order
approximations $ r_1 $, $ \theta_1 $ and quadratic terms combined using
one first-order -- $ r_1 $ or $ \theta_1 $ -- and one second-order
quantity -- $ r_2 $ or $ \theta_2 $. Since the second-order
terms must be linear combinations of $ \A_{\pm r} $
and $ \A_{\pm \theta} $, the governing equations take the form
\begin{equation}
\label{eq:res_3}
(\D{0}^2 + \omega_x^2) x_3 = - 2i\omega_x \D{2}\A_x  + 
2i\omega_x\D{2}\A_{-x}  +  \omega_x^2 \sum_{|\alpha|=3} C_\alpha^{(3,x)}
\A_r^{\alpha_1} \A_\theta^{\alpha_2} \A_{-r}^{\alpha_3}
\A_{-\theta}^{\alpha_4},
\end{equation}
where all constants $ C_\alpha^{(3,x)} $ are real. 

%
%

\begin{table}[t]
\begin{center}
\begin{tabular}{c@{~~}c@{~~}r}
\hline\rule{0ex}{3ex}\rule[-2.5ex]{0ex}{3ex}
$ \omega_\theta : \omega_r $  &
\textbf{Secular terms}  &
\textbf{Solvability condition~~~~~} \\
\hline\hline
\textrm{Outside}&\rule{0ex}{3ex}
$ \D{2}\A_r $, $ |A_r|^2 \A_r $, $ |A_\theta^2| \A_r $ &
$ 2i\D{2}\A_r - \omega_r \left[\kappa_r |A_r|^2 + \kappa_\theta |A_\theta|^2 \right]\A_r = 0 $
\\
\textrm{resonance}&\rule[-2.5ex]{0ex}{3ex}
$ \D{2}\A_\theta $, $ |A_r|^2 \A_\theta $, $ |A_\theta^2| \A_\theta $ &
$ 2i\D{2}\A_\theta - \omega_\theta \left[ \lambda_r |A_r|^2 + \lambda_\theta |A_\theta|^2 \right] \A_\theta = 0 $
\\
\hline\rule{0ex}{3ex}
$ 1:3 $ &
$ \A_\theta^3 $ &
$ 2i\D{2}\A_r - \omega_r \left[\kappa_r |A_r|^2 + \kappa_\theta |A_\theta|^2 \right]\A_r + \omega_r K \A_\theta^3 = 0 $
\\
&
$ \A_r \A_{-\theta}^2 $ &
$ 2i\D{2}\A_\theta - \omega_\theta \left[ \lambda_r |A_r|^2 + \lambda_\theta |A_\theta|^2 \right]
\A_\theta + \omega_\theta L\A_r \A_{-\theta}^2 = 0 $\rule[-2.5ex]{0ex}{3ex}
\\
\hline\rule{0ex}{3ex}
$ 1:1 $ &
$ |A_r|^2 \A_\theta $, $ |A_\theta|^2 \A_\theta $, &
$ 2i\D{2}\A_r - \omega_r \left[\kappa_r |A_r|^2 + \kappa_\theta |A_\theta|^2 \right]\A_r + \omega_r\times\qquad  $
\\
&
$ \A_r^2 \A_{-\theta} $, $ \A_{-r} \A_\theta^2 $ &\rule[-2ex]{0ex}{2ex}
$(K_1 |A_r|^2 \A_\theta + K_2 |A_\theta|^2 \A_\theta + K_3 \A_r^2 \A_{-\theta} + K_4 \A_{-r} \A_\theta^2) = 0 $
\\
&
$ |A_r|^2 \A_r $, $ |A_\theta|^2 \A_\theta $, &
$ 2i\D{2}\A_\theta - \omega_\theta \left[ \lambda_r |A_r|^2 + \lambda_\theta |A_\theta|^2 \right]
\A_\theta + \omega_\theta\times\qquad $
\\
&
$ \A_{-r} \A_\theta^2 $, $ \A_r^2 \A_{-\theta} $ &
$ (L_1 |A_r|^2 \A_r + L_2 |A_\theta|^2 \A_\theta + L_3 \A_{-r} \A_\theta^2 + L_4 \A_r^2 \A_{-\theta}) = 0 $\rule[-2.5ex]{0ex}{3ex}
\\
\hline\rule{0ex}{3ex}
$ 3:1 $ &
$ \A_{-r}^2 \A_\theta $ &
$ 2i\D{2}\A_r - \omega_r \left[\kappa_r |A_r|^2 + \kappa_\theta |A_\theta|^2 \right]\A_r + \omega_r K \A_{-r}^2 \A_\theta = 0 $
\\
&
$ \A_r^3 $ &
$ 2i\D{2}\A_\theta - \omega_\theta \left[ \lambda_r |A_r|^2 + \lambda_\theta |A_\theta|^2 \right]
\A_\theta + \omega_\theta L \A_r^3 = 0 $\rule[-2.5ex]{0ex}{3ex}
\\
\hline
\end{tabular}
\caption{Possible resonances in the third order of approximation.}
\label{tab:res_3}
\end{center}
\end{table}

The secular terms together with possible resonances are summarized in the table
\ref{tab:res_3}. Far from any resonance, we eliminate the terms which are secular
independently of $\omega_r$ and $\omega_\theta$. 
Multiplying by $ e^{-i \omega_x t} $, the
solvability conditions take the form
\begin{eqnarray}
\label{eq:res_3sec_r}
\D{2} A_r = - \frac{i \omega_r}{2} \left[ \kappa_r |A_r|^2 +
\kappa_\theta |A_\theta|^2 \right] A_r,
\\
\label{eq:res_3sec_m}
\D{2} A_\theta = - \frac{i \omega_\theta}{2} \left[ \lambda_r |A_r|^2 +
\lambda_\theta |A_\theta|^2 \right] A_\theta,
\end{eqnarray}
where we denoted $ \kappa_r = C^{(3,r)}_{2010} $, $ \kappa_\theta =
C^{(3,r)}_{1101} $ , $ \lambda_r = C^{(3,\theta)}_{1110} $ and $
\lambda_\theta = C^{(3,\theta)}_{0201} $ because of simpler notation. A
particular solution of equation (\ref{eq:res_3}) is given by linear
combination of cubic terms constructed from $ \A_{\pm r} $ and $ \A_{\pm
\theta} $
\begin{equation}
\label{eq:res_3sol}
x_3 = \sum_{|\alpha|=3} Q_\alpha^{(3,x)} \A_r^{\alpha_1}
\A_\theta^{\alpha_2} \A_{-r}^{\alpha_3} \A_{-\theta}^{\alpha_4},
\end{equation}
where all coefficients $ Q_\alpha^{(3,x)} $ are real.

The terms proportional to $ \epsilon^4 $ in the expanded left-hand side
of the equations (\ref{eq:res_gov_r}) and (\ref{eq:res_gov_theta}) are
\begin{equation}
\left[ \ddot{\delta x} + \omega_x^2 x \right]_4 = (\D{0}^2 + \omega_x^2)
x_3  +  2 \D{3}\D{0} x_1  + 2 \D{0} \D{2} x_2.
\end{equation}
The operator $ \D{0}\D{2} $ acts on $ x_2 $ given by
(\ref{eq:res_2sol}). The result is found using the solvability
conditions (\ref{eq:res_3sec_r}), (\ref{eq:res_3sec_m}) and can be
written in the form
\begin{equation}
\label{eq:res_3lhs}
2 \D{0} \D{2} x_2 = \omega_x^2 \sum_{|\alpha| = 4} J_\alpha^{(x)}
\A_r^{\alpha_1} \A_\theta^{\alpha_2} \A_{-r}^{\alpha_3}
\A_{-\theta}^{\alpha_4}.
\end{equation}
where constants $ J^{(x)}_\alpha $ are real because both $ \D{0} $ and $
\D{2} $ produce one ``$ i $''. The right-hand side is expanded similarly. 
We obtain
\begin{equation}
\label{eq:res_4}
(\D{0}^2 + \omega_x^2) x_4 = - 2i\omega_x \D{3}\A_x  + 
2i\omega_x\D{3}\A_{-x}  +  \omega_x^2 \sum_{|\alpha|=4} C_\alpha^{(4,x)}
\A_r^{\alpha_1} \A_\theta^{\alpha_2} \A_{-r}^{\alpha_3}
\A_{-\theta}^{\alpha_4},
\end{equation}
with real $ C_\alpha^{(4,x)} $. On the right-hand side there is only one
secular term $ - 2i\omega_x \D{3}\A_x $ independently of $\omega_r$ and $\omega_\theta$, the sum contains only
terms which becomes secular near a resonance. These terms
and solvability conditions are listed in the table \ref{tab:res_4}. 

%
%

\begin{table}[t]
\begin{center}
\begin{tabular}{c@{~~~}c@{~~~}r}
\hline
\rule{0ex}{3ex}\rule[-2.5ex]{0ex}{3ex}
$ \omega_\theta : \omega_r $  &
\textbf{Secular terms}  &
\textbf{Solvability condition~~} \\
\hline\hline
\textrm{Outside}&\rule{0ex}{3ex}
$ \D{3}\A_r $ &
$ \D{3}\A_r = 0 $
\\
\textrm{resonance}&$ \D{3}\A_\theta $ &
$ \D{3}\A_\theta = 0 $\rule[-2.5ex]{0ex}{3ex}
\\
\hline
\rule{0ex}{3ex}$ 1:4 $ &
$ \A_\theta^4 $ &
$ 2i\D{3}\A_r - \omega_r K \A_\theta^4 = 0 $
\\
&$ \A_r\A_\theta^3 $ &
$ 2i\D{3}\A_\theta - \omega_\theta L \A_r\A_\theta^3 = 0 $\rule[-2.5ex]{0ex}{3ex}
\\
\hline
\rule{0ex}{3ex}$ 2:3 $ &
$ \A_r \A_{-\theta}^3 $ &
$ 2i\D{3}\A_r - \omega_r K \A_r \A_{-\theta}^3 = 0 $
\\
&$ \A_r^2 \A_{-\theta}^2 $ &
$ 2i\D{3}\A_\theta - \omega_\theta L \A_r^2 \A_{-\theta}^2 = 0 $\rule[-2.5ex]{0ex}{3ex}
\\
\hline
\rule{0ex}{3ex}$ 3:2 $ &
$ \A_{-r}^2 \A_\theta^2 $ &
$ 2i\D{3}\A_r - \omega_r K \A_{-r}^2 \A_\theta^2 = 0 $
\\
&$ \A_r^3 \A_{-\theta} $&
$ 2i\D{3}\A_\theta - \omega_\theta L \A_r^3 \A_{-\theta} = 0 $\rule[-2.5ex]{0ex}{3ex}
\\
\hline
\rule{0ex}{3ex}$ 4:1 $ &
$ \A_r^3 \A_\theta $ &
$ 2i\D{3}\A_r - \omega_r K \A_r^3 \A_\theta = 0 $
\\
&$ A_r^4 $ &
$ 2i\D{3}\A_\theta - \omega_\theta L \A_r^4 = 0 $\rule[-2.5ex]{0ex}{3ex}
\\
\hline
\end{tabular}
\caption{Possible resonances in the fourth order of approximation.}
\label{tab:res_4}
\end{center}
\end{table}

One general feature of an internal resonance $ k:l $ is that $ k \omega_r
$ and $ l \omega_\theta $ need not to be infinitesimally close.
Consider, for example, resonance $ 1:2 $. The resonance occurs when $
\omega_\theta \approx 2 \omega_r $. Suppose that the system departs from
this exact ratio by small (first-order) deviation $ \omega _\theta = 2
\omega_r + \epsilon \sigma $, where $ \sigma $ is called detuning
parameter. Then the terms $ \A_{-r} \A_\theta $ and $ \A_r^2 $ in the
equations (\ref{eq:res_2}) remain still secular in $ T_0 $ since
\begin{equation}
\A_{-r}\A_\theta = A_r^\ast A_\theta e^{i (\omega_\theta - \omega_r)
T_0} = A_r^\ast A_\theta e^{i (\omega_r + \epsilon \sigma) T_0} =
A_r^\ast A_\theta e^{i \sigma T_1} e^{i \omega_r T_0}
\end{equation}
and analogically for $ \A_r^2 $. 

%
%

\section{The 3:2 parametric resonance}
\label{sec:32}

Let us study oscillations of the conservative system which
eigenfrequencies $ \omega_r $ and $\omega_\theta$ are close to $ 3:2 $ ratio. 
The time behavior of the observed frequencies 
$\obsom_r$ and $\obsom_\theta$ and amplitudes $a_r$ and $a_\theta$ 
of the oscillations
is given by the solvability conditions 
(\ref{eq:res_2sec}), (\ref{eq:res_3sec_r}) and (\ref{eq:res_3sec_m}). In
the fourth order we eliminate terms which become nearly secular.
For this purpose let us introduce detuning parameters $ \sigma_2 $ and
$ \sigma_3 $ according to
\begin{equation}
3 \omega_r = 2 \omega_\theta + \epsilon^2 \sigma_2 + \epsilon^3
\sigma_3,
\end{equation}
where the term $ \epsilon \sigma_1 $ is missing, because the complex
amplitude $ A $ depends only on time-scales $ T_2 $ and $ T_3 $. 
The secular terms in eq.\
(\ref{eq:res_4}) are eliminated if (see table~\ref{tab:res_4})
\begin{eqnarray}
\label{eq:32_3sec_r}
2i \D{3} A_r - \omega_r \alpha (A_r^2)^\ast A_\theta^2 e^{-i (\sigma_2
T_2 + \sigma_3 T_3)} &=& 0,
\\
\label{eq:32_3sec_theta}
2i \D{3} A_\theta - \omega_\theta \beta A_r^3 A_\theta^\ast e^{i
(\sigma_2 T_2 + \sigma_3 T_3)} &=& 0,
\end{eqnarray}
where $ \alpha $ and $ \beta $ are real constants depending on
properties of the system. Since $ A_r $ and $ A_\theta $ are complex,
the conditions (\ref{eq:32_3sec_r}) and (\ref{eq:32_3sec_theta})
together with (\ref{eq:res_2sec}) and (\ref{eq:res_3sec_r}) represents 8
real equations. This can be seen by substituting the polar forms $ A_r =
\frac{1}{2}\tilde{a}_r e^{i \phi_r} $ and $ A_\theta =
\frac{1}{2}\tilde{a}_\theta e^{i \phi_\theta} $, and by separating real and
imaginary parts. We obtain
\begin{eqnarray}
\label{eq:32_D2a_r}
\D{2} \tilde{a}_r &=& 0, 
\\
\label{eq:32_D2a_theta}
\D{2} \tilde{a}_\theta &=& 0,
\\
\label{eq:32_D2phi_r}
\D{2} \phi_r &=& -\frac{\omega_r}{8} \left[ \kappa_r \tilde{a}_r^2 +
\kappa_\theta \tilde{a}_\theta^2 \right],
\\
\label{eq:32_D2phi_theta}
\D{2} \phi_\theta &=& -\frac{\omega_\theta}{8} \left[ \lambda_r
\tilde{a}_r^2 + \lambda_\theta \tilde{a}_\theta^2 \right],
\\
\label{eq:32_D3a_r}
\D{3} \tilde{a}_r &=& \frac{\alpha \omega_r}{16} \tilde{a}_r^2
\tilde{a}_\theta^2 \sin (-3 \phi_r + 2 \phi_\theta - \sigma_2 T_2 -
\sigma_3 T_3),
\\
\label{eq:32_D3a_theta}
\D{3} \tilde{a}_\theta &=& \frac{\beta \omega_\theta}{16} \tilde{a}_r^3
\tilde{a}_\theta \sin (3 \phi_r - 2 \phi_\theta + \sigma_2 T_2 +
\sigma_3 T_3),
\\
\label{eq:32_D3phi_r}
\D{3} \phi_r &=& - \frac{\alpha \omega_r}{16} \tilde{a}_r
\tilde{a}_\theta^2 \cos (-3 \phi_r + 2 \phi_\theta - \sigma_2 T_2 -
\sigma_3 T_3),
\\
\label{eq:32_D3phi_theta}
\D{3} \phi_\theta &=& - \frac{\beta \omega_\theta}{16} \tilde{a}_r^3
\cos (3 \phi_r - 2 \phi_\theta + \sigma_2 T_2 + \sigma_3 T_3).
\end{eqnarray}
The amplitudes $ \tilde{a}_r $ and $ \tilde{a}_\theta $ of the
oscillations change slowly, because they depend only on $ T_3 $. Phases
$ \phi_r $ and $ \phi_\theta $ of oscillations are modified on both time
scales $ T_2 $ and $ T_3 $. The number of equations can be reduced by
introducing the phase function $ \gamma(T_2, T_3) = 2 \phi_\theta - 3
\phi_r - \sigma T_2 - \sigma_3 T_3 $. Then we get
\begin{eqnarray}
\D{3} \tilde{a}_r &=& \frac{\alpha \omega_r}{16} \tilde{a}_r^2
\tilde{a}_\theta^2 \sin \gamma,
\\
\D{3} \tilde{a}_\theta &=& - \frac{\beta \omega_\theta}{16}
\tilde{a}_r^3 \tilde{a}_\theta \sin \gamma,
\\
\D{2} \gamma &=& -\sigma_2 + \frac{\omega_\theta}{4} \left( \mu_r
\tilde{a}_r^2 + \mu_\theta \tilde{a}_\theta^2 \right),
\\
\D{3} \gamma &=& -\sigma_3 + \frac{\omega_\theta}{8} \tilde{a}_r \left(
\alpha \tilde{a}_\theta^2 - \beta \tilde{a}_r^2 \right) \cos \gamma,
\end{eqnarray}
were we used the fact that near the resonance $ \omega_r \approx (2/3)
\omega_\theta $ and we defined $ \mu_r = \kappa_r - \lambda_r $ and
$ \mu_\theta = \kappa_\theta - \lambda_\theta $. The situation can be
further simplified if we come back to unique physical time $ t $. Then
equations for evolution of $ \gamma $ are merged using $ d/dt =
\epsilon^2 \D{2} + \epsilon^3 \D{3}. $ We obtain
\begin{eqnarray}
\label{eq:32_r}
\dot{a}_r &=& \frac{\alpha \omega_r}{16} a_r^2 a_\theta^2 \sin \gamma,
\\
\label{eq:32_theta}
\dot{a}_\theta &=& - \frac{\beta \omega_\theta}{16} a_r^3 a_\theta \sin
\gamma,
\\
\label{eq:32_gamma}
\dot{\gamma} &=& - \sigma + \frac{\omega_\theta}{4} \left[\mu_r a_r^2 +
\mu_\theta a_\theta^2 + \frac{a_r}{2} \left( \alpha a_\theta^2 - \beta
a_r^2 \right) \cos \gamma \right],
\end{eqnarray}
where we defined $ a = \epsilon \tilde{a} $ and $ \sigma = \epsilon^2
\sigma_2 + \epsilon^3 \sigma^3 $.


\subsection{Steady-state solutions}

Steady-state solutions are characterized by constant amplitudes and frequencies of
oscillations. Such solutions represent singular points of the system governed by equations
(\ref{eq:32_r})--(\ref{eq:32_gamma}).

 It is obvious from equations
(\ref{eq:32_r}) and (\ref{eq:32_theta}) that the condition $ \dot{a}_r
= \dot{a}_\theta = 0 $ can be satisfied (with nonzero amplitudes) only if $
\sin \gamma = 0 $ (identically at all times), 
and thus also $ \dot{\gamma} = 0 $. In that case
equation (\ref{eq:32_gamma}) transforms to the algebraic equation
\begin{equation}
\frac{\sigma}{\omega_\theta} = \frac{1}{4} \left[\mu_r a_r^2 +
\mu_\theta a_\theta^2 \pm \frac{a_r}{2} \left( \alpha a_\theta^2 - \beta
a_r^2 \right)\right].
\end{equation}
The left-hand side can be expressed
using the eigenfrequency ratio $ R = \omega_\theta/\omega_r $ as
\begin{equation}
\frac{\sigma}{\omega_\theta} = -\frac{2}{R} \left( R - \frac{3}{2}
\right).
\end{equation}
Then we get
\begin{equation}
R = \frac{3}{2} - \frac{3}{16} \left( \mu_r a_r^2 + \mu_\theta
a_\theta^2 \right) \pm \frac{3}{32} a_r \left( \alpha a_\theta^2 - \beta
a_r^2 \right),
\end{equation}
were we neglected terms of order $ a^4 $. Note that the lowest
correction to eigenfrequencies is of order of $ a^2 $ -- for given
amplitudes $ a_r $, $ a_\theta $ steady-state oscillations occur when
the ratio of eigenfrequencies departs from $ 3/2 $ by deviation of
order of $ a^2 $.

The relation between observed frequencies of oscillations $ \obsom_r $,
$ \obsom_\theta $ and eigenfrequencies $ \omega_r $, $ \omega_\theta $
are given by the time derivative of phases $ \phi_r $ and $\phi_\theta$
\begin{eqnarray}
\omega_r^\star = \omega_r + \dot{\phi}_r,
&&
\omega_\theta^\star = \omega_\theta + \dot{\phi}_\theta.
\end{eqnarray}
We can find simple relation between observed frequencies and the
phase function
\begin{equation}
3\obsom_r - 2\obsom_\theta = 3\omega_r - 2\omega_\theta + (3\dot{\phi}_r
- 2\dot{\phi}_\theta) = \sigma + (3\dot{\phi}_r - 2\dot{\phi}_\theta) =
-\dot{\gamma}.
\end{equation}
For steady state solutions $ \dot{\gamma} = 0 $, and thus observed
frequencies are adjusted to exact $ 3:2 $ ratio even if eigenfrequencies
depart from it. 

Finally, let us derive explicit relations for $ \obsom_r $ and $
\obsom_\theta $ up to to the second order in amplitudes. Using the
equations (\ref{eq:32_D2phi_r}) and (\ref{eq:32_D2phi_theta}) we find
\begin{eqnarray}
\label{eq:32_obsom}
\obsom_r = \omega_r \left[ 1 - \frac{1}{8} \left( \kappa_r a_r^2 +
\kappa_\theta a_\theta^2 \right) \right],
&&
\obsom_\theta = \omega_\theta \left[ 1 - \frac{1}{8} \left( \lambda_r
a_r^2 + \lambda_\theta a_\theta^2 \right) \right].
\end{eqnarray}


\subsection{Integrals of motion}

\begin{figure}[t!]
\begin{center}
\includegraphics[scale=1.2]{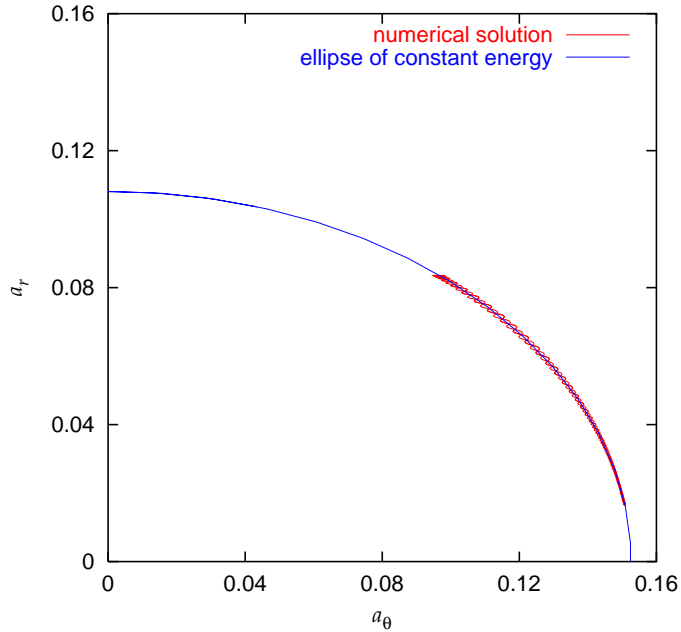}
\end{center}
\caption{Comparison between an analytical constraint (\ref{eq:32_E}) 
and the corresponding numerical solution of the system \citet{akklr03}. 
Each point
corresponds to the amplitudes of the oscillations at a particular time. On the
other hand, from the discussion of equation (\ref{eq:32_E}) we know that
these points must lay on an ellipse, whose shape is determined by the
multiple-scales method.}
\label{fig:32_ellipse}
\end{figure}

Behavior of the system is described by three variables $ a_r(t) $,
$ a_\theta(t) $ and $ \gamma(t) $  and three first-order differential
equations (\ref{eq:32_r}), (\ref{eq:32_theta}) and (\ref{eq:32_gamma}).
However, the number of differential equations can be reduced to one
because it is possible two find two integrals of motion of the system. 
Our discussion will be analogical to 
case of $1:2$ resonance of systems with quadratic nonlinearity,
as examined by \citet{nm79}.

Consider equations (\ref{eq:32_r}) and (\ref{eq:32_theta}). Eliminating
$ \sin \gamma $ from both equations we find
\begin{equation}
\frac{d}{dt}(a_r^2 + \nu a_\theta^2) = 0
\end{equation}
and thus
\begin{equation}
\label{eq:32_E}
a_r^2 + \nu a_\theta^2 = \mathrm{const} \equiv E,
\end{equation}
where we defined 
\begin{equation}
\label{eq:32_nu}
\nu = \frac{\alpha \omega_r}{\beta \omega_\theta} \approx \frac{2
\alpha}{3 \beta}.
\end{equation}
When $ \nu > 0 $, the both amplitudes of oscillations are bounded. The
curve $ [a_r(t), a_\theta(t)] $ is a segment of an ellipse. The constant
$ E $ is proportional to the energy of the system. On the other hand,
when $ \nu < 0 $, one amplitude of oscillations can grow without bounds
while the second amplitude vanishes. This case corresponds to the presence
of an regenerative element in the system \citep{nm79}. The corresponding
curve in the $(a_r,a_\theta)$ plane is a hyperbola. In further
discussion we assume that $ \nu > 0 $. 

In order to verify that the the energy of the system is conserved, we
numerically integrated governing equation (\ref{eq:res_gov_r}) and
(\ref{eq:res_gov_theta}) for the one particular system discussed by
\citet{akklr03}. The comparison is in figure \ref{fig:32_ellipse}. 
The numerical and analytical results are in very good agreement.

The second integral of motion is found in following way. Let us multiply
the equation (\ref{eq:32_gamma}) by $ a_\theta $. Then we obtain
\begin{equation}
a_\theta \dot{\gamma} = -\sigma a_\theta  +  \frac{\omega_\theta}{4}
\mu_r a_r^2 a_\theta  +  \frac{\omega_\theta}{4} \mu_\theta a_\theta^3 
+  \frac{\omega_\theta}{8} \alpha a_r a_\theta^3 \cos \gamma  - 
\frac{\omega_\theta}{8} \beta a_r^3 a_\theta \cos \gamma.
\end{equation}
Changing the independent variable from $ t $ to $ a_\theta $ and
multiplying the whole equation by $ d a_\theta $ we find
\begin{eqnarray}
\label{eq:32_derL1}
a_r^3 a_\theta^2 d(\cos \gamma)  +  \frac{8\sigma}{\beta \omega_\theta}
d(a_\theta^2)  -  \frac{4 \mu_r}{\beta} a_r^2 a_\theta d(a_\theta^2) -
\frac{\mu_\theta}{\beta} d(a_\theta^4) - 
\nonumber
\\
 - \frac{2\alpha}{\beta} a_r a_\theta^3 \cos \gamma d a_\theta  +  2
 a_r^3 a_\theta \cos \gamma d a_\theta = 0.
\end{eqnarray}
The equation (\ref{eq:32_E}) implies 
\begin{equation}
\label{eq:32_dE} a_\theta d a_\theta = -\frac{a_r d a_r}{\nu}. 
\end{equation}
With aid of this relation the equation (\ref{eq:32_derL1}) takes the
form
\begin{eqnarray}
3 a_r^2 a_\theta^2 \cos \gamma d a_r  +  2 a_r^3 a_\theta \cos \gamma d
a_\theta  +  a_r^3 a_\theta^2 d(\cos\gamma)  +   
\nonumber
\\
 + \frac{8\sigma}{\beta \omega_\theta} d(a_\theta^2) + 
\frac{\mu_r}{\beta \nu} d (a_r^4)  -  \frac{\mu_\theta}{\beta}
d(a_\theta^4)  = 0.
\end{eqnarray}
The first three terms express the total differential of function $ - a_r^3
a_\theta^2 \cos \gamma $. Hence, the above equation can be arranged
to the form
\begin{equation}
d \left( a_r^3 a_\theta^2 \cos \gamma  +  \frac{8\sigma}{\beta
\omega_\theta} a_\theta^2  +  \frac{\mu_r}{\beta \nu} a_r^4  - 
\frac{\mu_\theta}{\beta} a_\theta^4 \right) = 0.
\end{equation}
In other words,
\begin{equation}
\label{eq:32_L}
a_r^3 a_\theta^2 \cos \gamma  +  \frac{8\sigma}{\beta \omega_\theta}
a_\theta^2  +  \frac{\mu_r}{\beta \nu} a_r^4  - 
\frac{\mu_\theta}{\beta} a_\theta^4 = \mathrm{const} \equiv L
\end{equation}
is another integral of equations (\ref{eq:32_r})--(\ref{eq:32_gamma}). 


\subsection{Analytical results}

Knowing two integrals of motion, we should be able to find one differential equation
which governs behavior of the system.

First, the amplitudes $ a_r $ and $ a_\theta $ are not independent
because they are related by equation (\ref{eq:32_E}). To satisfy this
relation, let us define new variable $ \xi(t) $ by
\begin{eqnarray}
\label{eq:32_xi}
a_r^2 = \xi E,
&&
a_\theta^2 = (1-\xi)\frac{E}{\nu}.
\end{eqnarray}
The equation describing an evolution of $ \xi(t) $ is derived as
follows. Let us multiply equation (\ref{eq:32_r}) by $ 2 a_r $ and
integrate it. We obtain
\begin{equation}
\frac{d (a_r^2)}{dt} = \frac{\alpha}{8}\omega_r a_r^3 a_\theta^2 \sin
\gamma.
\end{equation}
Then we express $ a_r^2 $ using $ \xi $, and square it. We find
\begin{equation}
\label{eq:32_derxi}
\left(\frac{8 E}{\alpha \omega_r} \right)^2 \dot{\xi}^2 = \left( a_r^3
a_\theta^2 \sin \gamma \right)^2.
\end{equation}
The right-hand side of this equation can be expressed using
(\ref{eq:32_L})
\begin{equation}
\left( a_r^3 a_\theta^2 \sin \gamma \right)^2  =  \left( a_r^3
a_\theta^2 \right)^2  -  \left( L  -  \frac{8\sigma}{\beta
\omega_\theta} a_\theta^2  -  \frac{\mu_r}{\beta \nu} a_r^4  + 
\frac{\mu_\theta}{\beta} a_\theta^4 \right)^2.
\end{equation}

After the substitution into the equation (\ref{eq:32_derxi}) and using
the relations (\ref{eq:32_xi}), we get 
\begin{equation}
\label{eq:32_EOM}
\frac{1}{E^3} \left( \frac{8}{\beta\omega_\theta} \right)^2 \dot{\xi}^2 
=  (1-\xi)^2\xi^3  -  \frac{\nu^2}{E^5} \left[ L  -  \frac{8 \sigma
E}{\beta \nu \omega_\theta} (1-\xi)  -  \frac{\mu_r E^2}{\beta \nu}
\xi^2  +  \frac{\mu_\theta E^2}{\beta \nu^2} (1-\xi)^2 \right]^2.
\end{equation}
The equation of motion has very familiar form
\begin{equation}
\label{eq:32_EOM_form}
\mathcal{K}^2 \dot{\xi}^2 = F^2(\xi) - G^2(\xi),
\end{equation}
where the $ \mathcal{K}^2 $ is a positive constant, $ F(\xi) =
(1-\xi)\xi^{3/2} $ and $ G(\xi) $ is a quadratic function which
coefficients depend on initial condition through $ E $ and $ L $. For
example, the equation with an effective potential, which governs motion
of test particle around a massive body, has the same form.
Therefore the following discussion is identical as in that case.

\begin{figure}[t!]
\begin{center}
\includegraphics[scale=1.3]{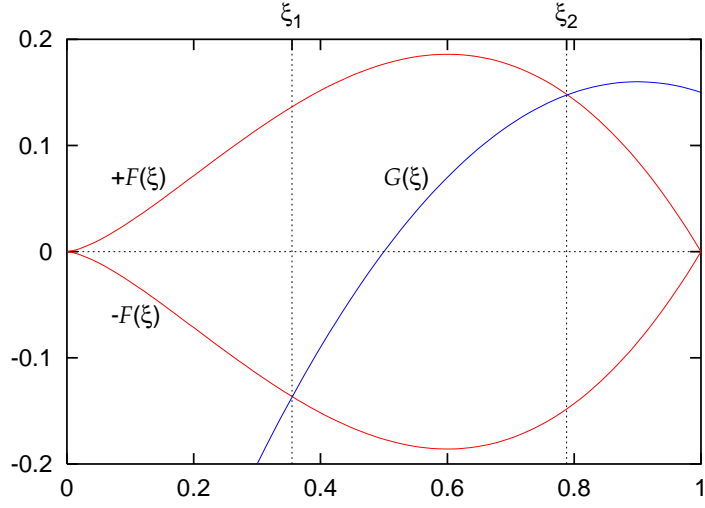}
\end{center}
\caption{The functions 
$ \pm F(\xi) = \pm (1-\xi)\xi^{3/2} $ and the
quadratic function $ G(\xi) $ which second powers are first and second terms on
the right-hand side of the equation (\ref{eq:32_EOM}). The behavior of
the system corresponds to $ \xi $ in the interval $ [\xi_1, \xi_2] $
(denoted by the two dotted vertical lines) where the condition $
|F(\xi)| \geq |G(\xi)| $ is satisfied. }
\label{fig:32_FG}
\end{figure}

In general, the motion occurs only when $ \dot{\xi}^2 $ is positive and
thus for $ \xi $ which satisfy $ |F(\xi)| \geq |G(\xi)| $. The turning
points, where $ \dot{\xi} $ changes its signature, are determined by the
condition
\begin{equation}
\label{eq:32_turn}
|F(\xi)| = |G(\xi)|.
\end{equation}

The functions $ \pm F(\xi) $ and $ G(\xi) $ are plotted in figure 
\ref{fig:32_FG}.  Generally, the function $ G $ intersects
functions $ \pm F $ in two points which corresponds to $ \xi(t) $
oscillating between two bounds $ \xi_1 $ and $ \xi_2 $ given by
condition (\ref{eq:32_turn}). In that case the radial and vertical mode
of oscillations will periodically exchange the energy. The exchanged
energy is given by $ \Delta E/E = \xi_2 - \xi_1 $. However, for some
particular values of $ L $ and $ E $ only one intersection of $ \pm F $
and $ G $ can be found. These stationary oscillations correspond to the
steady-state solutions discussed above.

\begin{figure}[t!]
\begin{center}
\includegraphics[width=0.6\textwidth]{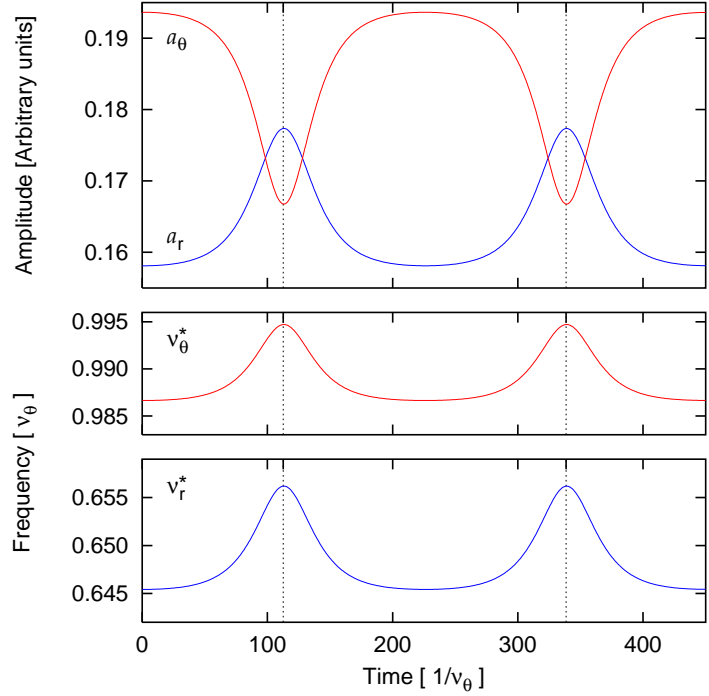}
\end{center}
\caption{Time evolution of the amplitudes (top panel) and 
the observed frequencies, $\nu_\theta^\star = \obsom_\theta/(2\pi)$
(middle panel) and
$\nu_r^\star = \obsom_r/(2\pi)$ (bottom panel). All quantities are
rescaled with respect the higher eigenfrequency $ \nu_\theta $. 
Amplitudes of the oscillations are anticorrelated
because the energy is conserved. Observed frequencies are correlated 
because the system is in parametric resonance.}
\label{fig:32_sol}
\end{figure}

The period of the energy exchange can be find by integration of the
equation (\ref{eq:32_EOM})
\begin{equation}
T = \frac{16}{\beta \omega_\theta} E^{-3/2} \int_{\xi_1}^{\xi_2} \frac{d
\xi}{\sqrt{F^2(\xi) - G^2(\xi)}}.
\end{equation}
The integral on the right-hand side can be estimated in the following
way. Since $ P_5 = F^2(\xi) - G^2(\xi) $ is a polynomial of the fifth
order in $ \xi $ having two roots $ \xi_1 $ and $ \xi_2 $ in the
interval $ [0,1] $, we can write it as $ -(\xi - \xi_1)(\xi - \xi_2)
P_3(\xi) $, where $ P_3(\xi)  $ is a polynomial of the third order
positive in the interval $ [0,1] $. Using the mean-value theorem we get 
\begin{equation}
\int_{\xi_1}^{\xi_2} \frac{d \xi}{\sqrt{-(\xi - \xi_1)(\xi - \xi_2)
P_3(\xi)}} = \frac{1}{p} \int_{\xi_1}^{\xi_2} \frac{d \xi}{\sqrt{-(\xi -
\xi_1)(\xi - \xi_2)}} = \frac{\pi}{p},
\end{equation}
where $ p > 0 $ is a value of $ P_3 $ for some $ \xi $ in the interval $
[\xi_1, \xi_2] $. Since $ P_5 \sim F^2 \sim 0.01 $ and $ (\xi_2 -
\xi_1)^2 \sim 0.01 $ typically, the values of $ P_3(\xi) $ are of order
of unity, and therefore $ p \sim 1 $. The period of the energy exchange can be
roughly approximated by
\begin{equation}
\label{eq:res32_T}
T \sim \frac{16 \pi}{\beta \omega_\theta} E^{-3/2}.
\end{equation}
However, near the steady state $ (\xi_2 - \xi_1)^2 $ approaches to zero and
the period becomes much longer.

The observed frequencies $ \obsom_r $ and $ \obsom_\theta $, given by
relations (\ref{eq:32_obsom}), depend on squares of amplitudes $ a_r $
and $ a_\theta $. Since both $ a_r^2 $ and $ a_\theta^2 $ depend linearly
on $ \xi(t) $, also observed frequencies are linear functions of
$ \xi $ and are linearly correlated. The slope of this
correlation $ \obsom_\theta = K \obsom_r + Q $ is independent of the
energy of oscillations and is given only by parameters of the system,
\begin{equation}
K = \frac{\omega_\theta}{\omega_r} \frac{\lambda_r \nu -
\lambda_\theta}{\kappa_r \nu - \kappa_\theta}.
\end{equation}
The slope of the correlation differs from $3:2$, however the observed
frequencies are still close to it.


\subsection{Numerical results}

The equations (\ref{eq:32_r})--(\ref{eq:32_gamma}) were solved
numerically using the fifth-order Runge-Kutta method with an adaptive step
size. One of the solutions is shown in figure \ref{fig:32_sol}. 
The top panel of the figure \ref{fig:32_sol} shows the time
behavior of the amplitudes of two modes of oscillations. Since energy of
the system is constant, amplitudes are anticorrelated and the two modes
are continuously exchanging energy between each other. 
The middle and the bottom panels show the two observed
frequencies that are mutually correlated. They are also correlated
to one of the amplitudes. The frequency ratio varies with time and it
differs from the exact $ 3:2 $ ratio, however, it always remains very close to it. 
The numerical solution is in agreement with the general results 
obtained analytically in the previous section. 

%
%

\section{Conclusions}

Although this paper was originally motivated by observations and
models connected to high-frequency QPOs, 
our results are very general and can be applied
to any system with governing equations of the form (\ref{eq:res_gov_r})
and (\ref{eq:res_gov_theta}). Moreover, the solvability conditions,
which are derived for all resonances up to the fourth order and summarized
in tables \ref{tab:res_2}, \ref{tab:res_3} and \ref{tab:res_4}, are
valid also for non-conservative systems. In the latter case, the 
only difference is that
constants $ C^{(n,x)}_\alpha $ that appear in the multiple scale
expansion are generally complex. However, the results discussed in 
section~\ref{sec:32} were derived under the assumption that the system is
conservative and thus all the constants $ C^{(n,x)}_\alpha $ are real.

The main result of this calculation is the prediction of low-frequency
modulation of the amplitudes and frequencies of oscillations. The
characteristic timescale is approximately given by equation
(\ref{eq:res32_T}). In a separate paper by \citet{h04} we pointed to possible
connection of this modulation with the \quot{normal branch oscillations}
(NBOs) that are often present together with QPOs. 
Specifically, we suggest that the correlation
between the higher frequency and the lower amplitude, evident in figure
\ref{fig:32_sol}, is the same as was recently reported in Sco X-1 by
\citet{ykj01}.

\bigskip

It is a pleasure to thank
Vladim{\'\i}r Karas, Marek Abramowicz, Wlodek Klu\'zniak,
Paola Rebusco, Michal Bursa and Michal Dov\v{c}iak for helpful
discussions. This work was supported by the GACR grant 205/03/0902 and
GAUK grant 299/2004.

%
%

\end{document}